\documentclass[jcp,groupedaddress,amssymb, amsmath]{revtex4-1}
\usepackage[version=3]{mhchem} 
\usepackage{longtable,array}
\usepackage{graphicx}
\usepackage{pbox}
\begin{document}
\title{Explicit treatment of hydrogen bonds in the Universal Force Field: validation and application for Metal-Organic Frameworks, hydrates and host-guest complexes}
\author{Damien E. Coupry}
\affiliation{Wilhelm-Ostwald-Institut f\"ur Physikalische und Theoretische Chemie, Fakult\"at f\"ur Chemie und Mineralogie, Universit\"at Leipzig, Linn\'estr. 2, 04103 Leipzig, Germany}
\author{Matthew A. Addicoat}
\email{matthew.addicoat@ntu.ac.uk}
\affiliation{School of Science and Technology, Nottingham Trent University, Clifton Lane, NG11 8NS Nottingham, UK}
\author{Thomas Heine}
\affiliation{Wilhelm-Ostwald-Institut f\"ur Physikalische und Theoretische Chemie, Fakult\"at f\"ur Chemie und Mineralogie, Universit\"at Leipzig, Linn\'estr. 2, 04103 Leipzig, Germany}


\begin{abstract}


A straightforward means to include explicit hydrogen bonds within the Universal Force Field is presented. Instead of treating hydrogen bonds as non-bonded interaction subjected to electrostatic and Lennard-Jones potentials, we introduce an explicit bond with negligible bond order, thus maintaining the structural integrity of the H-bonded complexes and avoiding the necessity to assign arbitrary charges to the system. The explicit hydrogen bond changes the coordination number of the acceptor site and the approach is thus most suitable for systems with under-coordinated atoms, such as many metal-organic frameworks, however, it also shows excellent performance for other systems involving a hydrogen-bonded framework. In particular, it is an excellent means for creating starting structures for molecular dynamics and for investigations employing more sophisticated methods.
 
The approach is validated for the hydrogen bonded complexes in the S22 dataset and then employed for a set of metal-organic frameworks from the Computation-Ready Experimental (CoRE) database and several hydrogen bonded crystals including water ice and clathrates. We show that direct inclusion of hydrogen bonds reduces the maximum error in predicted cell parameters from 66\% to only 14\% and the mean unsigned error is similarly reduced from 14\% to only 4\%. We posit that with the inclusion of hydrogen bonding, the solvent-mediated breathing of frameworks such as MIL-53 is now accessible to rapid UFF calculations, which will further the aim of rapid computational scanning of metal-organic frameworks while providing better starting points for electronic structure calculations.\end{abstract}

\maketitle
\section{Introduction}
\label{sec:intro}
Metal-Organic Frameworks (MOFs) are now a well-known class of crystalline, porous materials, where inorganic connectors are joined by organic linkers, forming a 3-dimensional network. Since the publication of MOF-5 in 1999\cite{Li_1999_Design-and-synthesis-of-an-exceptionally}, research into MOFs has grown almost exponentially and there are now several thousand synthesised MOFs.\cite{Chung_2014_Computation-Ready-Experimental-MetalOrganic} Several reviews have been written on various aspects of MOF chemistry,\cite{Odoh_2015_Quantum-Chemical-Characterization-of-the-Properties,Schneemann_2014_Flexible-metal-organic-frameworks} including their interactions with water.\cite{Burtch_2014_Water-Stability-and-Adsorption,Canivet_2014_Water-adsorption-in-MOFs:}\\

Water is an important solvent for MOFs, it serves as a synthetic medium,\cite{Choi_2004_Benzene-Templated-Hydrothermal-Synthesis,Yang_2007_Four-Novel-Three-Dimensional,Stock_2012_Synthesis-of-Metal-Organic-Frameworks} a structural component,\cite{DeCoste_2013_The-effect-of-water-adsorption} and an adsorbate.\cite{Furukawa_2014_Water-Adsorption-in-Porous} 
Water plays a major role in mediating defects and dissolution of many MOFs.\cite{St.-Petkov_2012_Defects-in-MOFs:-A-Thorough,Tan_2015_Water-interactions-in-metal,Kusgens_2009_Characterization-of-metal-organic-frameworks} Some MOFs, such as the UiO series, are highly water stable\cite{Cavka_2008_A-New-Zirconium-Inorganic-Building}, while others break down or transform\cite{Saha_2010_Structural-Stability-of-Metal,Greathouse_2006_The-Interaction-of-Water-with,Cheng_2009_Reversible-Structural-Change,Bezverkhyy_2014_MIL-53Al-under-reflux} when exposed to water. Proposed applications of MOFs include separation of organic contaminants\cite{Wang_2014_An-Exceptionally-Stable-and-Water-Resistant} and arsenic\cite{Vu_2015_Arsenic-removal-from} from aqueous mixtures and capture of water from air\cite{Furukawa_2014_Water-Adsorption-in-Porous}.\\

The requirement to computationally scan or screen the nearly infinite number of possible MOFs led to the development of UFF4MOF.\cite{Addicoat_2014_Extension-of-the-Universal-Force,Coupry_2016_Extension-of-the-Universal-Force} UFF4MOF extends the Universal Force Field of Rapp\'{e}\cite{Rappe_1992_UFF-a-full-periodic-table} to include several atom types present in MOFs but not accounted for in the original list of atom types. UFF4MOF thus permits the rapid calculation of structures for a wide range of actual and hypothetical MOFs. The structures predicted using UFF4MOF are typically within 5\% of experimental cell parameters, and it is therefore of interest whether similar accuracy can be maintained when the structural parameters are significantly determined by hydrogen bonding within the framework. MIL-53\cite{Loiseau_2004_A-Rationale-for-the-Large-Breathing} represents a notable case of water not only stabilizing, but determining the framework structure. The transition from (evacuated) large pore to narrow pore to hydrated and superhydrated large pore has been extensively studied by both experimental and computational methods.\cite{Salles_2011_Molecular-Insight-into,Cirera_2012_The-effects-of-electronic-polarization,Guillou_2011_Rapid-and-reversible-formation,Ortiz_2014_New-Insights-into-the-Hydrogen} The empty, large pore (LP) structures of MIL-53(M) (M = Cr, Al, Fe) have all been replicated using UFF4MOF, and in this work we seek to replicate the hydrated narrow pore (NP) structures, whose cell parameters are dictated by the hydrogen bonds that occur between the guest water molecules and the framework.\cite{Salazar_2015_Characterization-of-adsorbed-water,Ferey_2012_Swelling-Hybrid-Solids} \\


Treatment of hydrogen bonds is challenging for classical force fields. On one hand, the bonds are readily formed and broken in solution, so that an explicit, permanent bond would be an inappropriate description if investigating dynamical processes or, for example, the liquid state. Many force fields treat hydrogen bonds as non-bonded interactions: the bonding is achieved by attractive components arising from the London dispersion and the Coulomb interaction. In UFF, the former is defined by a Lennard-Jones potential, while the latter relies on the definition of charges. As the Coulomb interaction dominates, charge definition is crucial for the results. However, this approach works well for water and is particularly useful in QM/MM calculations, as was previously shown in DFTB/UFF calculations with electrostatic embedding.\cite{Heine_2007_Structure-and-Dynamics-of-b-Cyclodextrin,Lula_2007_Study-of-angiotensin-17-vasoactive}.\\

Several force fields attempt to explicitly treat hydrogen bonds: MM2\cite{Allinger_1977_Conformational-analysis.-130.}, MM3\cite{Allinger_1989_Molecular-mechanics.-The-MM3-force} and AMBER force fields all employ (optionally in the case of AMBER) an angle-independent 10,12 Lennard-Jones potential to describe nonbonded interactions. MM3 was later updated to include directional hydrogen bonding\cite{directional_MM3}. Paton and Goodman\cite{Paton_2009_Hydrogen-Bonding-and-p-Stacking} reviewed these force fields along with OPLS*, OPLSAA and MMFF, which do not include explicit hydrogen bonding and concluded that for the S22 database, the OPLSAA and MMFF force fields performed best, yielding accurate geometries and interaction energies. The same two force fields also ranked highest against the JSCH2005 database, but some optimized geometries differed from the benchmark geometry.\\

Despite these advances, there generally remains a technical problem. For host-guest systems that are treated entirely using a force field, it is often not practical to define charges to treat the electrostatic interactions. Indeed, although a charge equilibration scheme was designed for use with UFF \cite{doi:10.1021/j100161a070}, the original parameterization was done without an electrostatic model: the implementation is thus open to interpretation. Some codes forego explicit charges completely \cite{martin2013mcccs,O'Boyle2011}. UFF also explicitly includes electrostatic interactions for 1st and 2nd neighbours, so that most electrostatics is screened for a bonded system. This, of course, does not hold for non-bonded interactions, where significant electrostatic contributions arise due to the proximity of atoms formally carrying a charge. 
There is a further, merely technical aspect for the choice of explicit hydrogen bond that is crucial when studying static hydrogen-bonded frameworks including MOFs, ice, clathrates, or inclusion compounds: upon generation their starting structure, the non-bonded character in conjunction with substantial forces at initial geometry optimization, or, worse, kinetic energy gain upon molecular dynamics startup, makes it hard to converge to or to maintain the anticipated local minimum representing the desired framework topology.

\section{Definition of explicit hydrogen bonds within the Universal Force Field}
To provide a fast and topology-preserving structure generation method, and to avoid the definition of charges in order to account for the Coulomb interaction in the hydrogen bonds, we take advantage of the implicit definition of the electronegativity correction $r_{\textrm{EN}}$, which is contributing to the UFF natural bond length  
$r_{ij}$:
\begin{equation}
r_{ij} = r_i + r_j + r_{\textrm{BO}} - r_{\textrm{EN}} 
  \label{eqn:rij}
\end{equation}
where $r_i$ and $r_j$ are the bond radii of atoms $i$ and $j$ respectively, $r_{\textrm{BO}}$ is a bond order correction to the bond distance.
The bond order correction, in turn, is defined as:
\begin{equation}
r_{\textrm{BO}} = -\lambda(r_i + r_j) \ln(n)
  \label{eqn:rbo}
\end{equation}
where $\lambda$ is a proportionality constant derived using propane, propene and propyne with the C\_3, C\_2 and C\_1 radii and is equal to 0.1332. There is precedent for assigning unique bond orders for special types of bonds, with the amide bond order set to 1.41 in order to reproduce the C-N bond distance in $N$-methylformamide.\cite{Rappe_1992_UFF-a-full-periodic-table} Assigning a similar bond order for hydrogen bonds requires selecting an appropriate reference. At first glance, this is less than straightforward, given the variety of hydrogen bonds that may be encountered in framework structures. However, considering the prime importance of water to the synthesis, structure and behaviour of MOFs, a reasonable reference is the the prototypical $C_s$ global minimum of the water dimer. We further require that our treatment of hydrogen bonds can reproduce the hydrogen bond mediated breathing that occurs in hydrated MIL-53.

Treating hydrogen bonds as explicit bonds without bond order does have negative consequences: The dynamical breaking and formation of hydrogen bonded networks, essential for the description of the liquid state, becomes impossible (though it may work if the Reactive Force Field, ReaxFF\cite{Duin_2001_ReaxFF:-A-Reactive-Force}, was used instead of UFF). We note that defining an explicit hydrogen bond increases the coordination number of the central atom by one, however, this typically does not pose a problem for otherwise under-coordinated sites, and fully coordinated atoms typically are not hosting extra solvent. Secondly, because the defined bonding network is changed in the hydrogen-bonded system, it is not possible to calculate binding energies or relative stabilities using this approach. However, for the problem of rapid structural pre-optimization, neither of these disadvantages apply and furthermore, they are offset by the ability to rapidly produce high quality structures, without requiring the relatively expensive calculation of electrostatic terms. 


As we will show below, for ``frozen" configurations of hydrogen bonded dimers our approach performs significantly better than the traditional UFF treatment using nonbonded interactions with Coulomb interactions defined via HF/3-21G charges, and should provide a much better starting point for subsequent electronic structure calculations.\\

The water dimer has been extensively studied over decades, by both experimental\cite{Bouteiller_2011_The-vibrational-spectrum-of-the-water,Serov_2014_Rotationally-resolved-water} and computational\cite{Song_2014_Characteristics-of-hydrogen-bond,Wang_2014_Ultrafast-nonadiabatic-dynamics,Ronca_2014_A-Quantitative-View-of-Charge} means. Early calculations yielded a H$\cdots$O distance of 1.72~\AA\cite{Morokuma_1968_MolecularOrbital-Studies-of-Hydrogen}, a subsequent study including some of the same authors yielded a value of 1.80~\AA\cite{Morokuma_1970_Molecular-Orbital-Studies}. By the 1990's, calculations using MP2 and Coupled Electron Pair Theory (CEPA-1)\cite{Duijneveldtvan-de-Rijdt_1992_Convergence-to-the-basisset-limit} yielded geometries and energies in good agreement with experiment\cite{Odutola_1980_Partially-deuterated-water,Dyke_1977_The-structure-of-water-dimer} and the ``gold standard" CCSD(T)/QZ geometry found in the S22 database\cite{Jurecka_2006_Benchmark-database-of-accurate} also agrees very well, with $R_{H\cdots O} = 1.952 \AA$ (corresponding to $R_{OO} = 2.91 \AA$). If we therefore consider the UFF bond length and bond order correction in Equations \ref{eqn:rij} and \ref{eqn:rbo}, we note that $\lambda$, $r_i$ and $r_j$ and $ r_{\textrm{EN}}$ are all fixed within the UFF framework. Therefore, for water, where the oxygen atom  has the O\_3 atom type and hydrogen has the H\_ atom type, $r_i + r_j = 0.354 + 0.658 = 1.012\AA$, the electronegativity correction for a H-O\_3 bond is 0.0021 and thus the required bond order correction is equal to 0.9380~\AA. A bond order of 0.001 yields correction of 0.9312~\AA\ and noting the diminishing returns of further optimization of such a ``bond order", combined with the large range of acceptable bond lengths for a hydrogen bond, we propose to this value as a reasonable approximation to a bond order for describing hydrogen bonds in UFF calculations on metal-organic frameworks. Employing this bond order yields a bond length of 2.025~\AA\ for a H$\cdots$N\_R bond and 1.899~\AA\ where an O\_2 atom is the proton acceptor, both of which are reasonable lengths for a hydrogen bond.\\ 

\section{Results and Discussion}
\label{sec:rd}
To check the basic sensibility of using a bond order to correct hydrogen bond distances, we undertook geometry optimizations of the hydrogen bonded complexes in the S22 database. The $C_{2h}$  ammonia dimer was excluded from analysis as the angle terms resulting from the addition of the two hydrogen bonds considerably change the geometry of the dimer. The hydrogen bond distances of the other six complexes are shown in Table \ref{tab:S22_hbonds} and this is the only metric we employ for these non-bonded clusters.\cite{Witte_2015_Beyond-Energies:-Geometries}  Calculations were undertaken in the General Utility Lattice Program (GULP)\cite{Gale_2003_The-General-Utility-Lattice,Gale_2005_GULP:-Capabilities-and-prospects} except those employing atomic charges, where deMonNano\cite{deMonNano} was employed. Hartree-Fock calculations were undertaken in Gaussian09\cite{g09}. \\ %

\begin{table}
  \caption{\ Hydrogen bond distances calculated using UFF for hydrogen bonded complexes in the S22 database\protect{\cite{Jurecka_2006_Benchmark-database-of-accurate}}. Percent errors are calculated as ($X_{\textrm{UFF}}$ - $X_{\textrm{ref}}$)/$X_{\textrm{ref}} \times 100$, where $X_{\textrm{UFF}}$ denotes UFF-predicted and $X_{\textrm{ref}}$ denotes the original \emph{ab initio} value.}
  \label{tab:S22_hbonds}
  \begin{tabular}{llllll}
    \hline\\

    Complex (symmetry) & \pbox{4cm}{Hydrogen bond\\ Atom types} & \pbox{3cm}{UFF with\\ HF/3-21G \\charges} &\pbox{3cm}{UFF with\\ explicit\\ H-bonds} & Reference& \% error \\
    \hline\\
    \ce{(H2O)2} ($C_s$) 				& O\_3-H$\cdots$O\_3	& 2.666	& 1.920	& 1.952  	& -1.7	\\	
    ammonia dimer ($C_{2h}$)			& N\_3-H$\cdots$N\_3	& 2.854	& -		& 2.504	& -		\\
    Formic acid dimer ($C_{2h}$)		& O\_R-H$\cdots$O\_1	& 2.488	& 1.886	& 1.670	& 12.9	\\	
    Formamide dimer ($C_{2h}$)		& N\_R-H$\cdots$O\_1	& 2.520	& 1.884	& 1.840	& 2.4		\\	
    Uracil dimer ($C_{2h}$)				& N\_R-H$\cdots$O\_2	& 2.481	& 1.882	& 1.774	& 6.1		\\	
    2-pyridoxine - 2-aminopyridine ($C_1$)	& N\_R-H$\cdots$N\_R	& 2.609	& 2.030	& 1.860	& 9.1		\\	
    								& N\_R-H$\cdots$O\_1 	& 2.585	& 1.858	& 1.874	& -0.8	\\	
    Adenine - thymine WC ($C_1$)		& N\_R-H$\cdots$N\_R 	& 2.556	& 2.030	& 1.819	& 11.6	\\	
    								& N\_R-H$\cdots$O\_1 	& 2.477	& 1.857	& 1.929	& -3.7	\\	
    
    \hline
  \end{tabular}
\end{table}

\begin{table}
  \caption{\ Error on bond angles calculated using UFF for hydrogen bonded complexes in the S22 database\protect{\cite{Jurecka_2006_Benchmark-database-of-accurate}}. Percent errors are calculated as $100/N_{\textrm{ang}}\times\sum\limits_{i=1}^{N_{\textrm{ang}}}{(X_{\textrm{UFF}} - X_{\textrm{ref}})/X_{\textrm{ref}}}$ and rmse as 
$(1/N_{\textrm{ang}}\times\sum\limits_{i=1}^{N_{\textrm{ang}}}{(X_{\textrm{UFF}} - X_{\textrm{ref}})^2})^{\frac{1}{2}}$, 
where $X_{\textrm{UFF}}$ 
denotes UFF-predicted and $X_{\textrm{ref}}$ denotes the original \emph{ab initio} value}
  \label{tab:S22_hbonds_angles}
\begin{tabular}{lcccc}
\hline\\
{} &      \multicolumn{2}{c}{UFF, LJ only}         & \multicolumn{2}{c}{UFF with explicit H-bonds}                     \\
{Complex (symmetry)} &     RMSE & Mean error (\%)&            RMSE & Mean error (\%)\\
\hline\\
\ce{(H2O)2} ($C_s$)                         &  2.97341 &              2.6813 &         3.60928 &             2.85337 \\
ammonia dimer ($C_{2h}$)                        &  6.89258 &             7.00757 &         22.4794 &             21.6555 \\
Formic acid dimer ($C_{2h}$)                    &  6.97037 &             22.1015 &         7.08619 &             81.4685 \\
Formamide dimer ($C_{2h}$)                      &  9.29674 &             39.9108 &         2.85878 &             14.4514 \\
Uracil dimer ($C_{2h}$)                &  3.60883 &             9.89631 &         3.02618 &             9.90769 \\
2-pyridoxine - 2-aminopyridine ($C_1$) &  3.57988 &             16.3517 &         2.85318 &             10.1884 \\
Adenine - thymine WC ($C_1$) &   3.7847 &             15.6545 &         4.11144 &             11.9441 \\
\hline
\end{tabular}
\end{table}

For these simple complexes, the results show reasonable agreement with the \emph{ab initio} reference and are far superior to UFF + Lennard-Jones + electrostatics. The two  N\_R-H$\cdots$N\_R bonds are overestimated by approximately 10\%, as is the unusually short hydrogen bond in the formic acid dimer. Other bonds are within 6\% of their reference values. Undertaking the calculations without specification of the hydrogen bond results in bond distances increasing by approximately 1\AA\, which is clearly poor.\\  
We note in Table \ref{tab:S22_hbonds}, the only failure of the explicit hydrogen-bond approach is the $C_{2h}$ ammonia dimer, which upon inclusion of two explicit hydrogen bonds, optimizes to a singly hydrogen-bonded complex of $C_{s}$ symmetry. In this case the two hydrogen bonds in the reference structure form H-N$\cdots$H angles of 58$^\circ$, a significant deviation from the 106.7$^\circ$ angle of the N\_3 parameter, and the optimizer prefers to allow a single hydrogen bond with a close-to-ideal angle (i.e. essentially tetrahedral geometry around the acceptor nitrogen atom), rather than two hydrogen bonds with large errors on the angle term. This is a general limitation of this approach, in that by making the hydrogen bond explicit, the coordination number of both the hydrogen atom and the accepting atom are increased by one and the hydrogen bonded atom consequently figures in the angle terms around the acceptor atom. In most cases, this effect is either desired or benign, such as in the case of hydrogen bonding to an under-coordinated metal atom in a paddlewheel. \\


\subsection{Metal-Organic Frameworks}
Having thus established the validity of the approach, a set of framework materials where hydrogen bonding is important for maintaining structural integrity was selected from the Computation Ready Experimental (CoRE) database.\cite{Chung_2014_Computation-Ready-Experimental-MetalOrganic} The original crystal structures were re-sourced from the Cambridge Structural Database\cite{Allen_2002_The-Cambridge-Structural-Database} in order to recover the solvent molecules. To this test set we add the particular case of MIL-53(Al) NP\cite{Loiseau_2004_A-Rationale-for-the-Large-Breathing}. After assigning atom types to each structure, we detect hydrogen bonds by looking for hydrogen atoms. For every H atom that is directly connected to (O, S, N) we assign a larger covalent radius (of 2\AA), then with this, we re-build the connectivity list and we add to the original connectivity list every new bond between these super-big hydrogen atoms and (O, N, S, F, Cl, Br, I) provided that the bond angle is between 140 and 220 degrees. For all structures we calculate the structure both with and without specification of hydrogen bonds and note the resultant cell parameters, which are listed in Table \ref{tab:CoRE_hbonds}.\\

{
\setlength{\extrarowheight}{-5pt}
\begin{longtable}{llrrrrr}
\caption{Comparison of UFF calculated and experimental cell parameters of selected MOFs. The first 10 structures possess primarily intermolecular hydrogen bonds, the second 10 possess more intramolecular hydrogen bonds. Percent errors are calculated as ($X_{\textrm{UFF}}$ - $X_{\textrm{exp}}$)/$X_{\textrm{exp}} \times 100$, where $X_{\textrm{UFF}}$ denotes UFF-predicted and $X_{\textrm{exp}}$ denotes the original value.}
  \label{tab:CoRE_hbonds}\\
    \hline
    CSD Refcode & & Experimental & \pbox{3cm}{UFF without\\ H-bonds} & \pbox{3cm}{UFF with\\ H-bonds}&\pbox{3cm}{ \% error without\\ H-bonds} & \pbox{3cm}{\% error with\\ H-bonds}\\
    \hline

MIL-53(Al) NP\cite{Loiseau_2004_A-Rationale-for-the-Large-Breathing}    &\emph{a} = &19.504	&18.593	&19.423	&-4.7	&-0.4\\
    &\emph{b} =     & 15.201	&21.839	&15.427	&43.7	&1.5\\
    &\emph{c} =     &6.569	&6.366	&6.470	&-3.1	&-1.5\\
CDLGLU01\cite{CDLGLU01}    &\emph{a} =     &11.575 &10.099 &11.217 &-12.8  &-3.1\\
    &\emph{b} =     &10.764 &15.304 &9.920  &42.2   &-7.8\\
    &\emph{c} =     &7.256  &7.435  &7.775  &2.5    &7.2\\
CUGLTM\cite{CUGLTM}  &\emph{a} =     &11.084 &13.404 &10.989 &20.9   &-0.9\\
    &\emph{b} =     &10.350 &11.262 &9.412  &8.8    &-9.1\\
    &\emph{c} =     &7.238  &6.330  &7.246  &-12.6  &0.1\\
FUFREE\cite{FUFREE}  &\emph{a} =     &7.785  &8.312  &7.808  &6.8    &0.3\\
    &\emph{b} =     &10.238 &12.172 &9.780  &18.9   &-4.5\\
    &\emph{c} =     &15.851 &15.524 &16.074 &-2.1   &1.4\\
HURNOX\cite{HURNOX}  &\emph{a} =     &7.149  &6.891  &6.158  &-3.6   &-13.9\\
    &\emph{b} =     &10.468 &14.307 &9.573  &36.7   &-8.5\\
    &\emph{c} =     &11.295 &10.156 &11.518 &-10.1  &2.0\\
ICOWON\cite{ICOWON}  &\emph{a} =     &8.754  &8.240  &8.232  &-5.9   &-6.0\\
    &\emph{b} =     &10.003 &10.693 &9.775  &6.9    &-2.3\\
    &\emph{c} =     &11.790 &11.446 &11.479 &-2.9   &-2.6\\
JUCXEK\cite{JUCXEK}  &\emph{a} =     &11.382 &13.971 &12.306 &22.7   &8.1\\
    &\emph{b} =     &11.382 &14.020 &12.509 &23.2   &9.9\\
    &\emph{c} =     &11.734 &9.592  &10.169 &-18.3  &-13.3\\
MAZTIR\cite{MAZTIR}  &\emph{a} =     &19.014 &19.839 &19.215 &4.3    &1.1\\
    &\emph{b} =     &6.843  &9.715  &6.782  &42.0   &-0.9\\
    &\emph{c} =     &9.420  &9.217  &9.753  &-2.2   &3.5\\
MEHPAQ\cite{MEHPAQ}  &\emph{a} =     &7.221  &6.350  &7.205  &-12.1  &-0.2\\
    &\emph{b} =     &10.295 &11.218 &9.948  &9.0    &-3.4\\
    &\emph{c} =     &11.041 &13.425 &10.966 &21.6   &-0.7\\
MUTVUT\cite{MUTVUT}  &\emph{a} =     &9.335  &9.255  &9.399  &-0.9   &0.7\\
    &\emph{b} =     &22.203 &25.664 &23.168 &15.6   &4.3\\
    &\emph{c} =     &27.514 &26.756 &27.710 &-2.8   &0.7\\
SALLAT\cite{SALLAT}  &\emph{a} =     &15.686 &12.126 &15.023 &-22.7  &-4.2\\
    &\emph{b} =     &8.165  &13.522 &8.865  &65.6   &8.6\\
    &\emph{c} =     &13.119 &12.225 &12.767 &-6.8   &-2.7\\
    \hline
DUQSEO\cite{DUQSEO}  &\emph{a} =     &8.884  &7.930  &7.606  &-10.7  &-14.4\\
    &\emph{b} =     &13.093 &11.963 &12.266 &-8.6   &-6.3\\
    &\emph{c} =     &13.135 &16.660 &14.021 &26.8   &6.7\\
FAPTUN\cite{FAPTUN}  &\emph{a} =     &14.304 &11.704 &14.757 &-18.2  &3.2\\
    &\emph{b} =     &16.970 &17.091 &16.907 &0.7    &-0.4\\
    &\emph{c} =     &11.098 &11.342 &11.373 &2.2    &2.5\\
KOJCUI\cite{KOJCUI}  &\emph{a} =     &18.326 &21.776 &18.356 &18.8   &0.2\\
    &\emph{b} =     &25.300 &21.273 &24.369 &-15.9  &-3.7\\
    &\emph{c} =     &7.524  &6.804  &6.900  &-9.6   &-8.3\\
RATVEP\cite{RATVEP}  &\emph{a} =     &7.764  &9.142  &8.044  &17.8   &3.6\\
    &\emph{b} =     &10.177 &11.938 &10.240 &17.3   &0.6\\
    &\emph{c} =     &15.949 &15.840 &16.222 &-0.7   &1.7\\
SARBOE\cite{SARBOE}  &\emph{a} =     &14.389 &14.925 &14.477 &3.7    &0.6\\
    &\emph{b} =     &15.475 &14.511 &15.006 &-6.2   &-3.0\\
    &\emph{c} =     &8.424  &10.610 &8.788  &26.0   &4.3\\
SIVKAK\cite{SIVKAK}  &\emph{a} =     &7.583  &9.243  &8.340  &21.9   &10.0\\
    &\emph{b} =     &31.086 &27.033 &29.315 &-13.0  &-5.7\\
    &\emph{c} =     &13.799 &14.199 &13.523 &2.9    &-2.0\\
VEFLUP\cite{VEFLUP}  &\emph{a} = &8.002  &7.283  &7.460  &-9.0   &-6.8\\
    &\emph{b} =     &9.091  &10.877 &9.031  &19.6   &-0.7\\
    &\emph{c} =     &12.935 &13.070 &13.071 &1.0    &1.0\\
YORZAH\cite{YORZAH}  &\emph{a} =     &16.493 &16.283 &16.334 &-1.3   &-1.0\\
    &\emph{b} =     &6.820  &7.905  &6.597  &15.9   &-3.3\\
    &\emph{c} =     &22.220 &23.331 &21.906 &5.0    &-1.4\\
YUVSUE\cite{YUVSUE}  &\emph{a} =     &15.435 &17.188 &15.518 &11.4   &0.5\\
    &\emph{b} =     &15.435 &17.188 &15.793 &11.4   &2.3\\
    &\emph{c} =     &22.775 &18.122 &21.005 &-20.4  &-7.8\\
ZNGLUD\cite{ZNGLUD}  &\emph{a} =     &11.190 &13.840 &12.126 &23.7   &8.4\\
    &\emph{b} =     &10.463 &11.273 &10.112 &7.7    &-3.4\\
    &\emph{c} =     &7.220  &6.369  &7.203  &-11.8  &-0.2\\    
    \hline    
 & & & & Maximum Unsigned Error &65.6	&14.4\\
 & & & & Mean Unsigned Error &13.9 &	4.0\\
    
\end{longtable}

}

In framework systems, the explicit inclusion of hydrogen bonds should yield even better agreement with experimental reference structure, as terms corresponding to regular covalent bonds far outnumber the contribution due to hydrogen bonds, and this is indeed the case. Without specifying hydrogen bonds, each structure has at least one cell parameter expand by 15\% or greater, and the mean unsigned error on all cell parameters is 13.9\%. Once hydrogen bonds are specified, the maximum unsigned error in cell parameters is only 14.4\% and the mean unsigned error is only 4.0\%.\\
 
\begin{figure}
\centering
\includegraphics[width=15cm]{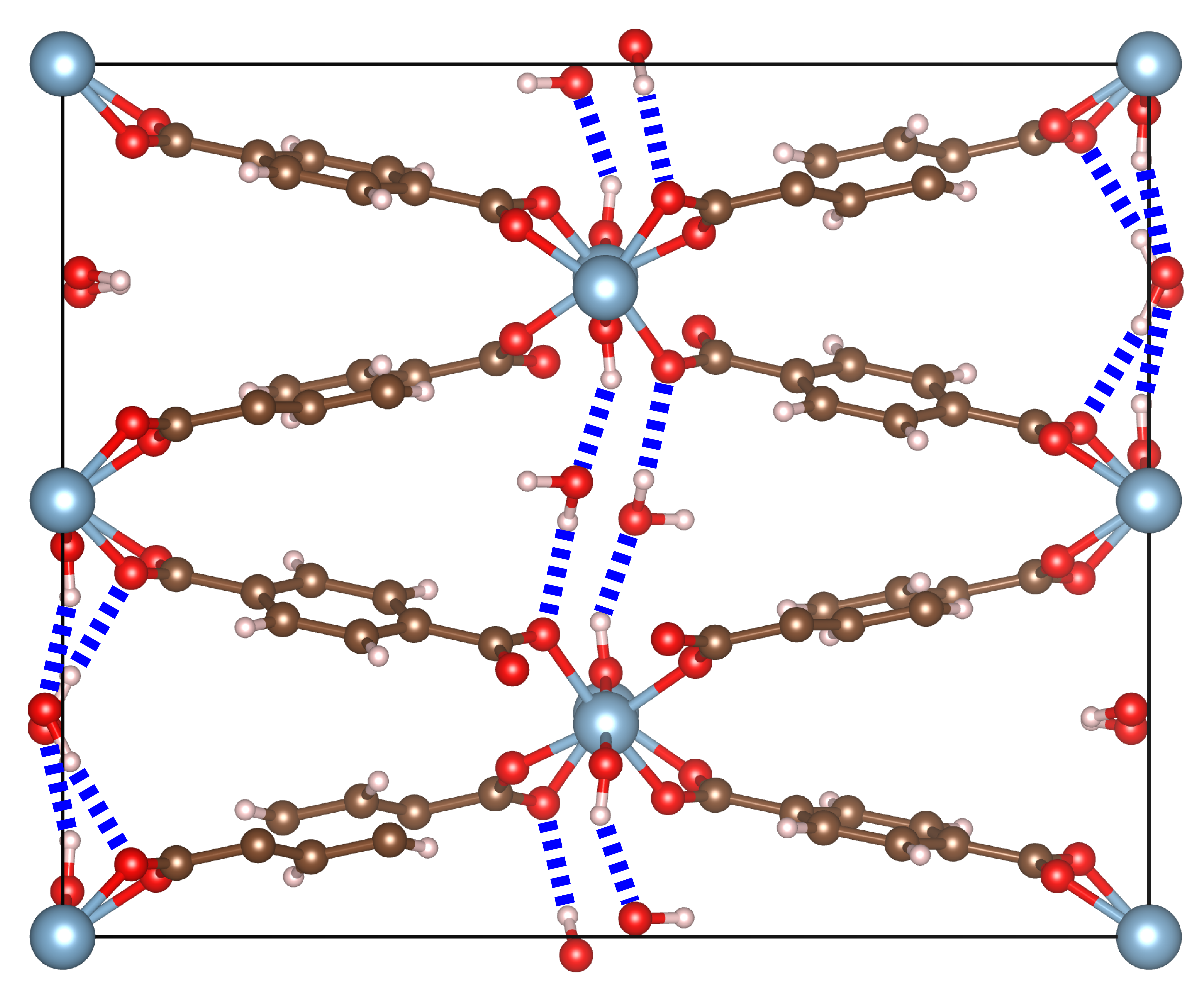}
\caption{MIL-53(Al) narrow pore structure (CCSD refcode: SABWAU
\protect{\cite{Loiseau_2004_A-Rationale-for-the-Large-Breathing}}) with hydrogen bonds between included water and framework indicated by dashed blue lines.}
\label{fig:SABWAU}
\end{figure}

The narrow pore structure of MIL-53(Al) is of particular note. Each pore contains two water molecules which are hydrogen-bonded to the framework oxygen atoms and the hydrogen of the framework hydroxy group. Optimising the structure with 16 hydrogen bonds specified (two hydrogen bonds per water molecule, see Figure \ref{fig:SABWAU}), the cell parameters of MIL-53(Al) are all predicted within 1.5\%, including the $b$ dimension, which otherwise expands by over 43\% to resemble the large pore structure.\\

\subsection{Hydrogen-bonded crystal structures}
The prediction of the structures of hydrogen-bonded MOFs is perhaps the primary use-case of this approach, however, the structure of any static hydrogen-bonded system is amenable to calculation this way. To illustrate the approach' broader utility, a variety of structures where hydrogen bonding is crucial to defining the structure were recalculated. Hydrogen bonds were inserted following the literature description of each structure. Table \ref{tab:ice_hbonds} shows the results for these structures.\\ 
\subsubsection{Methanol}
\begin{figure}
\centering
\includegraphics[width=15cm]{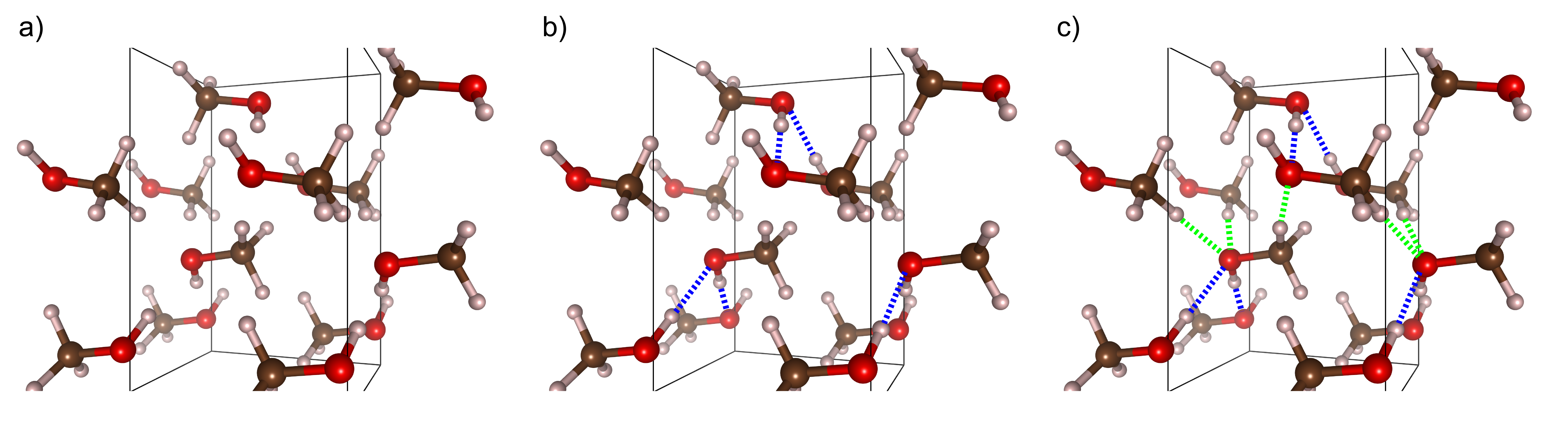}
\caption{$\alpha$-methanol with a) no explicit hydrogen bonds, b) explicit O-H$\cdots$O hydrogen bonds only and c) adding C-H$\cdots$O hydrogen bonds to b). A single unit cell contains four methanol molecules.}
\label{fig:aMeOH}
\end{figure}

The case of $\alpha$-methanol\cite{aMeOH} deserves special consideration as it illustrates how the choice of hydrogen bonds affects the optimized structure. There are two distinct types of hydrogen bond in $\alpha$-methanol; a bond from the hydroxyl hydrogen to the neighbouring oxygen with a OH$\cdots$O distance of 1.75\AA\ and a CH$\cdots$O distance of 2.51\AA, both illustrated in Figure \ref{fig:aMeOH}. Choosing only the strong OH$\cdots$O results in a correct estimate of the $a$ parameter, but exchanges the errors on the $b$ and $c$ parameters. Including the CH$\cdots$O in addition results in an underestimation of both the $b$ and $c$ parameters and a moderate overestimation of $a$. While none of the three calculated structures is a clear winner, arguably including only the OH$\cdots$O results in the best quality structure. This choice is then consistent with the structure of $\beta$-methanol\cite{bMeOH}, the high temperature phase, which possesses only OH$\cdots$O hydrogen bonds. These bonds, arranged in sheets, are crucial to the accurate description of the structure, neglecting them results in a strong distortion of the crystal, with the $a$ parameter being underestimated by 32\% and the $c$ parameter being overestimated by 48\%.\\
\subsubsection{Water ice and hydrate structures}
The classic hydrogen-bonded structure is that of water. The approach of fixing hydrogen bonds as actual bonds, makes the description of fluxional systems, such as liquid water, impossible. However, the structure of non-fluxional ice is readily amenable to calculation, the structures of both cubic and hexagonal water ice were calculated using all hydrogen bonds, and making each oxygen atom formally tetrahedral with two single bonds and two hydrogen bonds. Employing this approach results in a nearly uniform overestimation of lattice parameters of 4.5-7.5\% for both structures, a clear improvement over neglecting the hydrogen bonds which results in parameters being over/underestimated by 20-40\%.\\
The success of the explicit approach to hydrogen bonds in water ice has further significance in describing clathrate structures. Methane hydrates, in particular are the subject of renewed research as, found on the ocean floor and in polar regions they are estimated to contain up to 12\% of all the organic carbon on Earth\cite{Gbaruko2007192}, making them an important energy resource. In addition, hydrates are often formed within gas pipelines, where they are unwanted and cause significant damage\cite{Sloan_2003_Fundamental-principles-and-applications}. The structures of three methane hydrates, MH-I, MH-II and MH-H (hexagonal)\cite{jp111328v} were calculated, yielding results broadly similar to those of water ice, whereby specification of hydrogen bonds gives a structure with cell parameters uniformly overestimated by approximately 7\% and neglecting those bonds results in very poor structures with parameters under- and overestimated by up to 40\%. More modest improvement is seen for the structure of $n$-butanol hexahydrate (CCSD Refcode WUVZIW).\\ 
\subsubsection{Host-guest inclusion complexes and cocrystals}
To further illustrate the diversity of hydrogen-bonded systems to which this simple approach may be applied, structures described as being hydrogen-bonded were sourced from the CCSD\cite{Allen_2002_The-Cambridge-Structural-Database}. Structures included two urea-based inclusion compounds (ABAZOS\cite{ABAZOS} and WARWOB\cite{WARWOB}), a macrocyclic inclusion complex, ABUCIJ\cite{ABUCIJ}, a porous diamide matrix, ABEBUF\cite{ABEBUF}, a Cu coordination compound forming a hydrogen-bonded helicate, SIYRAU\cite{SIYRAU} and three cocrystals of pyrogallol[4]arenes and the ionic liquid 1-ethyl-3-methylimidazolium ethylsulfate\cite{DIZDOH}. In these cases, neglecting hydrogen bonds typically leads to at least one cell parameter in error by greater than 10\%. Overall, employing explicit hydrogen bonds reduces the maximum unsigned error for these complexes from 48.3\% to 12.1\% and the mean unsigned error from 12.6\% to 4.2\%.\\

{
\begin{table}
\scriptsize
  \caption{\ Comparison of UFF calculated and experimental cell parameters of selected hydrogen bonded crystals. Percent errors are calculated as ($X_{\textrm{UFF}}$ - $X_{\textrm{exp}}$)/$X_{\textrm{exp}} \times 100$, where $X_{\textrm{UFF}}$ denotes UFF-predicted and $X_{\textrm{exp}}$ denotes the original value.}
  \label{tab:ice_hbonds}
  \begin{tabular}{llrrrrr}
    \hline\\
    CSD Refcode & & Experimental & \pbox{3cm}{UFF without\\ H-bonds} & \pbox{3cm}{UFF with\\ H-bonds}&\pbox{3cm}{ \% error without\\ H-bonds} & \pbox{3cm}{\% error with\\ H-bonds}\\
    \hline\\

Ic ice\cite{ice}	&\emph{a}	&6.358	&8.805	&6.839	&38.5	&7.6\\
	&\emph{b}	&6.358	&9.019	&6.641	&41.8	&4.5\\
	&\emph{c}	&6.358	&9.019	&6.641	&41.8	&4.5\\
Ih ice\cite{ice}	&\emph{a}	&4.506	&5.483	&4.756	&21.7	&5.5\\
	&\emph{b}	&4.506	&6.284	&4.717	&39.5	&4.7\\
	&\emph{c}	&7.346	&5.205	&7.776	&-29.2	&5.9\\
alpha methanol\cite{aMeOH}	&\emph{a}	&4.873	&4.645	&4.955	&-4.7	&1.7\\
Figure \ref{fig:aMeOH}(c)	&\emph{b}	&4.641	&5.000	&4.487	&7.7	&-3.3\\
	&\emph{c}	&8.867	&9.835	&7.798	&10.9	&-12.1\\
alpha methanol\cite{aMeOH}	&	&	&	&4.876	&	&0.1\\
Figure \ref{fig:aMeOH}(b)	&	&	&	&5.178	&	&11.6\\
	&	&	&	&8.592	&	&-3.1\\
beta methanol\cite{bMeOH}	&\emph{a}	&6.409	&4.362	&6.702	&-31.9	&4.6\\
	&\emph{b}	&7.199	&7.268	&7.099	&1.0	&-1.4\\
	&\emph{c}	&4.649	&6.895	&4.554	&48.3	&-2.0\\
ammonia\cite{NH3}	&a	&5.138	&5.313	&4.919	&3.4	&-4.3\\
	&\emph{b}	&5.138	&5.313	&4.919	&3.4	&-4.3\\
	&\emph{c}	&5.138	&5.313	&4.919	&3.4	&-4.3\\
Methane Hydrate-I\cite{jp111328v}	&\emph{a}	&11.620	&12.181	&12.492	&4.8	&7.5\\
	&\emph{b}	&11.620	&16.164	&12.490	&39.1	&7.5\\
	&\emph{c}	&11.620	&10.436	&12.431	&-10.2	&7.0\\
Methane Hydrate-II\cite{jp111328v}&\emph{a}	&11.890	&12.262	&12.745	&3.1	&7.2\\
	&\emph{b}	&11.890	&11.775	&12.723	&-1.0	&7.0\\
	&\emph{c}	&11.890	&15.180	&12.712	&27.7	&6.9\\
Methane Hydrate-H\cite{jp111328v}	&\emph{a}	&11.910	&10.826	&12.789	&-9.1	&7.4\\
	&\emph{b}	&11.910	&14.074	&12.813	&18.2	&7.6\\
	&\emph{c}	&9.894	&11.302	&10.456	&14.2	&5.7\\
WUVZIW\cite{WUVZIW}	&\emph{a}	&7.400	&7.393	&7.355	&-0.1	&-0.6\\
	&\emph{b}	&24.448	&24.958	&25.248	&2.1	&3.3\\
	&\emph{c}	&14.265	&15.399	&14.707	&8.0	&3.1\\
PgC$_2$ Cocrystal 5 \cite{DIZDOH}	&\emph{a}	&10.059	&9.642	&9.693	&-4.1	&-3.6\\
	&\emph{b}	&14.556	&15.927	&13.886	&9.4	&-4.6\\
	&\emph{c}	&15.558	&17.731	&14.945	&14.0	&-3.9\\
PgC$_4$ Cocrystal 6\cite{DIZDOH}&\emph{a}	&21.075	&23.775	&21.334	&12.8	&1.2\\
	&\emph{b}	&12.796	&13.363	&12.452	&4.4	&-2.7\\
	&\emph{c}	&18.229	&18.353	&17.438	&0.7	&-4.3\\
PgC$_4$ Cocrystal 7\cite{DIZDOH}&\emph{a}	&21.075	&23.490	&20.088	&11.5	&-4.7\\
	&\emph{b}	&12.796	&13.506	&12.380	&5.6	&-3.2\\
	&\emph{c}	&18.229	&18.251	&17.118	&0.1	&-6.1\\
ABAZOS\cite{ABAZOS}	&\emph{a}	&16.338	&17.022	&16.306	&4.2	&-0.2\\
	&\emph{b}	&21.933	&22.805	&20.839	&4.0	&-5.0\\
	&\emph{c}	&16.338	&16.870	&16.310	&3.3	&-0.2\\
ABEBUF\cite{ABEBUF}	&\emph{a}	&10.722	&10.989	&10.851	&2.5	&1.2\\
	&\emph{b}	&10.900	&11.831	&11.230	&8.5	&3.0\\
	&\emph{c}	&27.635	&25.745	&27.648	&-6.8	&0.0\\
ABUCIJ\cite{ABUCIJ}	&\emph{a}	&11.281	&10.924	&11.076	&-3.2	&-1.8\\ 
	&\emph{b}	&17.888	&18.370	&17.973	&2.7	&0.5\\
	&\emph{c}	&23.950	&24.319	&23.863	&1.5	&-0.4\\
SIYRAU\cite{SIYRAU}	&\emph{a}	&13.456	&13.861	&13.257	&3.0	&-1.5\\
	&\emph{b}	&14.394	&13.727	&14.647	&-4.6	&1.8\\
	&\emph{c}	&15.935	&18.019	&16.945	&13.1	&6.3\\
WARWOB\cite{WARWOB}	&\emph{a}	&19.297	&22.634	&20.706	&17.3	&7.3\\
	&\emph{b}	&4.616	&5.599	&4.589	&21.3	&-0.6\\
	&\emph{c}	&8.705	&6.954	&7.972	&-20.1	&-8.4\\

\hline
 & & & & Maximum Unsigned Error &48.3	&12.1\\
 & & & & Mean Unsigned Error &12.6 &	4.2\\

  \end{tabular}
\end{table}

}

\section{Potential Applications}

When used to predict solvent effects, the accuracy of the structures produced by our approach is highly dependant on the guessed positions of the solvent molecules. For the structures where case by case reasoning is impossible, using a genetic algorithm (e.g: as implemented in the Atomic Simulation Environment\cite{ISI:000175131400009,doi:10.1063/1.4886337}) to maximize the number of hydrogen bonds in the structure of interest can yield reasonable starting points, enabling the use of explicit bonding in automated tasks. 
An active area of research like water confinement in carbon nanotubes of varying diameters \cite{Pascal19072011}, where the hydrogen bonding is the single most important factor for structure determination, would likely benefit from cheap, systematic structure elucidation. Appropriate modifications to the ASE code are underway to allow for solvent rigid motions during optimizations, and a proof of concept script is available on github (https://github.com/DCoupry/GenAlgHbond). 
However, the main application of explicit hydrogen bonding is case by case reasoning for a reasonable pre-optimization, followed by electronic structure methods. More complex properties of the structures, like frequencies (for which UFF was not parametrized) or the effects of H-bond anisotropy, fall squarely outside of the scope of this paper.

\section{Conclusions}
We report a simple approach to explicitly treat hydrogen bonds within the Universal Force Field. The approach does not require any definition of atomic charges, which is a great technical (codes such as GULP and ADF do not include long-range and non-bonded electrostatic interactions within UFF) and practical (the definition of atomic charges is not well-defined) advantage. Moreover, it avoids the computationally costly electrostatic interaction term and is equally applicable for periodic and non-periodic systems. 

The specification of hydrogen bonds with negligible (0.001) bond order increases the coordination number of the acceptor site by one, which is compatible with the definition of the main acceptor sites O and N, as well as with under-coordinated metal sites as present in metal-organic frameworks.

The approach was validated for the hydrogen bonded complexes in the S22 database. We show that it is very effective for both framework-framework hydrogen bonds and importantly, for framework-adsorbate bonds. Using this approach, both the large pore and narrow pore structures of MIL-53 can be calculated accurately with UFF. We show the generality of the approach by also applying it to hydrogen bonded crystal structures and host-guest inclusion complexes, including the environmentally and commercially important clathrates,

Finally, we note that this approach does not require any implementation, and thus works in any software that includes a UFF implementation. We propose this method to be most useful for pre-optimization and screening of static hydrogen-bonded systems.\\


\section{Acknowledgements}
This work was partially supported by the European Commission (ERC StG C3ENV (GA 256962), MSC-EID PROPAGATE (316897), DFG FOR 2433 FlexMOF (HE 3543/31-1) and the VolkswagenStiftung.

\newpage
\renewcommand \bibname{References}
\bibliography{UFF4MOF-hbonds_refs}

\end{document}